%% file: main.tex
\newif\ifblindreview
\blindreviewfalse 

\ifblindreview
\documentclass[sigconf,authordraft,9pt]{acmart}
\else
\documentclass[sigconf,nonacm,9pt]{acmart}
\fi
\AtBeginDocument{%
  }

\acmConference[Conference acronymous submission]{Blind Rev}{Earth}{YYYY}

\usepackage{algorithm}
\usepackage{algpseudocode}
\usepackage{booktabs} 
\usepackage{graphicx} 
\usepackage[table]{xcolor} 

\usepackage{amssymb} 
\usepackage{enumitem}
\usepackage{multirow}

\begin{document}
\settopmatter{printacmref=False}

\title{Scalable IP Mimicry: End-to-End Deceptive IP Blending to Overcome Rectification and Scale Limitations of IP Camouflage}


\ifblindreview
\author{This part is removed for blind review \vspace{1.2cm}}
\else
\author{\normalsize{Junling Fan\textsuperscript{1}, 
George Rushevich\textsuperscript{1}, 
Giorgio Rusconi\textsuperscript{1}, 
Mengdi Zhu\textsuperscript{2}, 
Reiner Dizon-Paradis\textsuperscript{1}, 
Domenic Forte\textsuperscript{1}} \\
{\textit{Department of Electrical and Computer Engineering\textsuperscript{1}, University of Florida}, \\
\textit{Department of Psychology\textsuperscript{2}, University of Florida} \\
\{fan.j, rushevich.g, grusconi, zhum, reinerdizon\}@ufl.edu, dforte@ece.ufl.edu}
}

\fi

\ifblindreview
\renewcommand{\shortauthors}{This part is removed for blind review}
\else
\renewcommand{\shortauthors}{Junling Fan, George Rushevich, Giorgio Rusconi, Mengdi Zhu, Reiner Dizon-Paradis, Domenic Forte}
\fi

\begin{abstract}
Semiconductor intellectual property (IP) theft incurs estimated annual losses ranging from \$225 billion to \$600 billion. 
Despite initiatives like the CHIPS Act, many semiconductor designs remain vulnerable to reverse engineering (RE). 
IP Camouflage is a recent breakthrough that expands beyond the logic gate hiding of traditional camouflage through ``mimetic deception,'' where an entire module masquerades as a different IP. However, it faces key limitations: requires a high-overhead post-generation rectification step, is not easily scalable, and uses an AIG logic representation that is mismatched with standard RE analysis flows.
This paper addresses these shortcommings by introducing two novel, end-to-end models. We propose a Graph-Matching algorithm to solve the representation problem and a DNAS-based NAND Array model to achieve scalability. To facilitate this, we also introduce a mimicry-aware partitioning method, enabling a divide-and-conquer approach for large-scale designs. Our results demonstrate that these models are resilient to SAT and GNN-RE attacks, providing efficient and scalable paths for end-to-end deceptive IP design.
\end{abstract}

\maketitle

\input{Section1_Introduction}

\input{Section3_BackgroundAndRelatedWork}

\input{Section4_MimicryAwarePartitioning}
\input{Section5_GraphMatching}

\input{Section6_NANDGateArray}

\input{Section7_ExperimentAndAnalysis}

\input{Section8_Conclusion}
\newpage

\bibliographystyle{ACM-Reference-Format}
\bibliography{references.bib}

\end{document}

%% file: Section1_Introduction.tex
\section{Introduction}

Integrated Circuits (ICs) are susceptible to a range of threats, including intellectual property (IP) theft through reverse engineering (RE), malicious modifications like the insertion of hardware Trojans, and unauthorized overproduction by untrusted entities in the supply chain. These threats not only cause billions in financial losses for semiconductor companies~\cite{arancaGlobalResearch} but also pose a critical risk to national security and critical infrastructure that rely on the integrity of electronic hardware.
To counter these threats, a variety of hardware security techniques have been developed. Among the most promising is IC camouflaging, a physical design strategy aimed at thwarting RE attacks at the layout level. The fundamental principle of IC camouflaging~\cite{rajendran2013security,erbagci2016secure,malik2015development} is to make different logic gates, such as NAND and NOR, physically indistinguishable during RE. By eliminating the unique visual signatures of standard cells, an attacker who has delayered a chip for analysis would be unable to definitively determine the function of each gate, making the reconstruction of the circuit's netlist computationally intractable.
However, traditional camouflaging methods are often applied as a post-design step, which can lead to significant overheads in power, performance, and area (PPA)~\cite{shakya_covert_2019,fan2025designing}. 

A more effective strategy is to integrate security considerations into the early stages of design. The state of the art in this domain, ``mimetic deception''~\cite{fan2025designing}, aims to make an IP block mimic a different IP block. Although this methodology effectively deceives ML-based RE methods~\cite{alrahis2021gnn}, it has critical flaws: its reliance on post-generation rectification compromises security and incurs more overhead than necessary, and it is limited to small logic cones. Furthermore, it relies on an And-Inverter Graph (AIG) representation. While AIGs are efficient for AI-based circuit encoding and model training, they do not reflect the standard logic gates (e.g., NAND, NOR, XOR, etc.) found in process design kits (PDKs), increasing gate count and logic depth and reducing both area and performance.
Further, since existing RE tools are optimized for standard logic-gate designs, they provide limited insight into how effective the deception based on AIGs would be in practice.

This paper addresses these limitations by introducing and exploring two new, end-to-end models that eliminate the need for post-generation rectification and present a clear trade-off between logic representation and scalability: (1) \textbf{Graph-Matching} addresses the representation problem. It models the circuit using standard logic gates, which directly mirrors the ``real logic'' representation that both design tools and attackers actually encounter; (2) \textbf{DNAS-based NAND Gate Array} solves the end-to-end and scale problem, allowing for the generation of large-scale deceptive circuits, but, like AIG-based methods, is limited to a single gate (NAND) representation.
Furthermore, to achieve the scalability for both models, we introduce a \textbf{mimicry-aware partitioning method} and deception-oriented matching process that strategically decomposes large circuits into smaller, manageable sub-problems. This enables the generation methods to be applied effectively to large-scale designs. We comprehensively explore these new methods and their inherent trade-offs. Our analysis reveals that the Graph Matching method is superior for ensuring formal correctness and standard gate representation, while the DNAS NAND Array excels in scalability, PPA efficiency, and achieving the highest overall mimicry.


The remainder of this paper is structured as follows: Section~\ref{sec:back} details the threat model that motivates this work, as well as background on camouflaging.
The methodologies for mimicry-aware partitioning, graph-matching, and DNAS-based NAND Gate Array are discussed in Sections~\ref{sec:partition} and \ref{sec:Method}.
Section~\ref{sec:results} describes our experiments for evaluating the proposed methods, comparing them to each other and the current state of the art. 
Conclusions and directions for future work are found in Section~\ref{sec:conclusion}.

%% file: Section3_BackgroundAndRelatedWork.tex
\section{Background and Motivation} \label{sec:back}

\subsection{Invasive IC RE and Understanding}

\begin{figure}[t]
    \centering
    \includegraphics[width=\columnwidth]{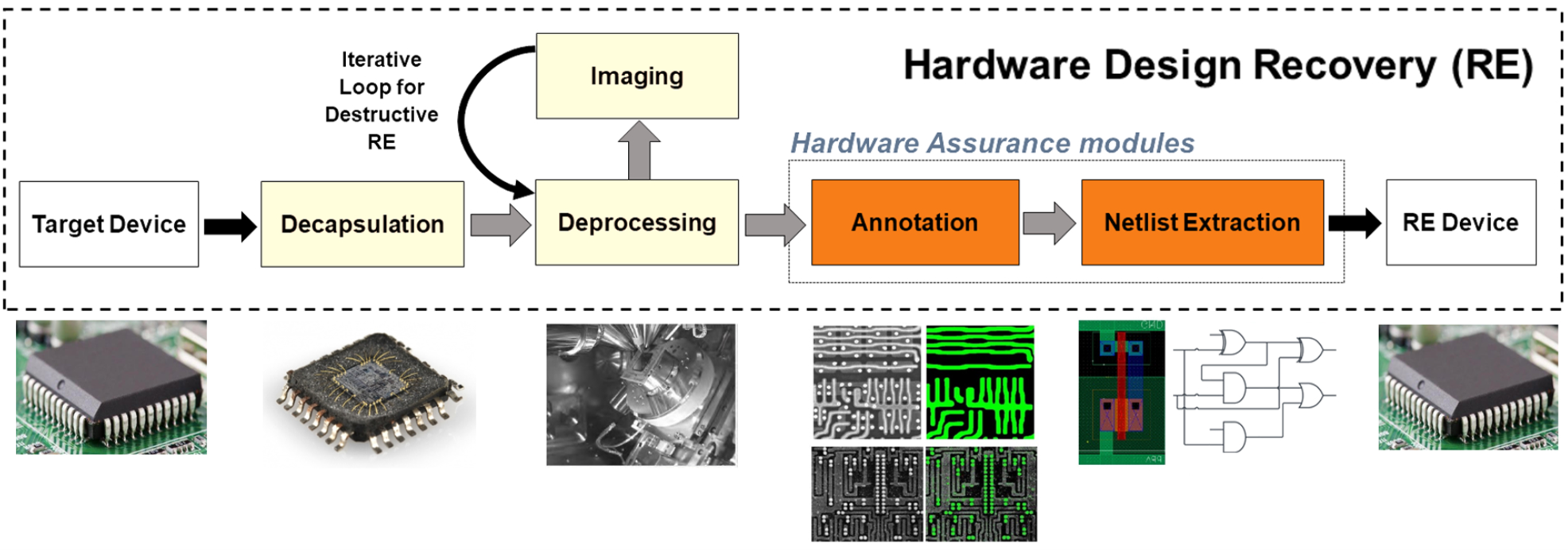}
    \caption{Hardware Design Recovery (RE) workflow.}
    \label{fig:RE}
    
\end{figure}

Physically invasive reverse engineering (RE) is a destructive process where an attacker systematically dissects an IC to uncover its complete netlist~\cite{torrance2011state,quadir2016survey}. The typical workflow, as illustrated in Figure \ref{fig:RE}, begins with decapsulation to expose the silicon die. Then, the attacker performs iterative deprocessing, where each metal and dielectric layer of the chip is carefully removed, often through a combination of mechanical polishing and advanced, automated ion beam milling~\cite{principe2017steps}. After each layer is stripped, high-resolution Scanning Electron Microscope (SEM) imaging is used to capture the circuit layout. These images are analyzed to annotate components and connections, ultimately allowing for the full netlist extraction of the design~\cite{quadir2016survey}. 
If the IC is unprotected, this netlist reveals the true intellectual property (IP). If camouflaged, the netlist is obscured, becoming the starting point for further logical analysis and attacks.

Next, netlist understanding abstracts high-level functions from a gate-level ``sea of gates''~\cite{albartus2020dana}. Since flattened netlists lack hierarchy and labels, automated analysis is essential for reconstructing meaningful design structures~\cite{subramanyan2013reverse, albartus2020dana}.
Algorithmic and formal approaches rely on structural and functional analysis~\cite{subramanyan2013reverse}. They identify recurring patterns (e.g., bitslices), infer datapaths, or extract control logic such as FSMs through Boolean reasoning~\cite{meade2016netlist}. Dataflow-based tools like DANA recover register structures by tracking signal dependencies~\cite{albartus2020dana}. These techniques are effective but often domain-specific and computationally heavy. 
Recently, machine learning-based approaches have leveraged Graph Neural Networks (GNNs) to interpret circuits as graphs and classify gate functions~\cite{alrahis2021gnn}. GNN-RE learns structural and functional boundaries directly from data, while FGNN2 introduces self-supervised contrastive pretraining to capture intrinsic logic functionality, recognizing equivalent implementations despite structural variation~\cite{wang2024fgnn2}.

\subsection{IC Camouflaging}
\begin{figure*}[t]
    \centering
    \includegraphics[width=0.99\textwidth]{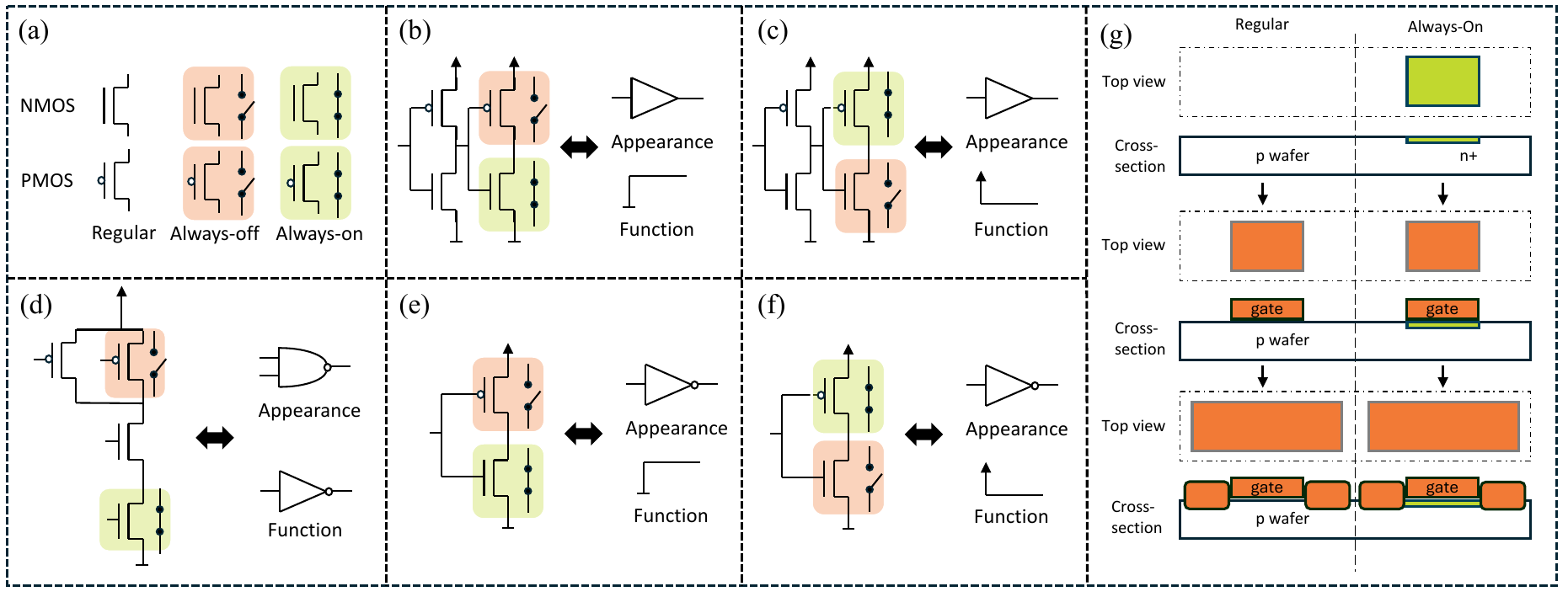}
    \vspace{-10pt}
    \caption{The covert gate~\cite{shakya_covert_2019} cells used in IP Camouflage~\cite{fan2025designing} to create function-appearance mismatches. (a)Regular, always-off, and always-on NMOS and PMOS transistors. (b) A Fake Buffer (FB) that appears as a buffer but functions as a constant logic 0. (c) An FB that appears as a buffer but functions as a constant logic 1. (d) A Fake NAND gate that appears as a NAND but functions as an inverter. (e) A Fake Inverter (FI) that appears as an inverter but functions as a logic 0. (f) An FI that appears as an inverter but functions as a logic 1. (g) An example manufacturing cross-section illustrating how an always-on transistor is created.}
    \label{fig:CovertGate}
    
\end{figure*}
IC camouflaging techniques are generally divided into two main categories~\cite{shakya_covert_2019}: gate-level and interconnect-level camouflaging.
Gate camouflaging replaces conventional logic gates with variants capable of realizing multiple functions depending on fabrication-dependent parameters, such as dummy contacts~\cite{rajendran2013security}, threshold voltage variations~\cite{erbagci2016secure}, or selective doping~\cite{malik2015development}.
This approach conceals the true functionality of gates from reverse engineering by SEM imaging.
In contrast, interconnect camouflaging~\cite{yu2017incremental} manipulates wiring connections between gates, obscuring the circuit’s signal paths and logic flow, thereby impeding accurate reconstruction of the netlist during IC reverse engineering.

To the best of our knowledge, the current state-of-the-art in IC camouflaging is the covert gate methodology~\cite{shakya_covert_2019} because it has never been successfully reverse engineered in the literature.
Covert gates integrate both gate and interconnect camouflage principles by exploiting always-on and always-off transistors~\cite{shakya_covert_2019}.
By controlling these transistor states during doping, designers can realize multiple gate functionalities and introduce dummy inputs or connections within the same physical layout.
Because these transistor states are indistinguishable under SEM, covert gates appear identical to standard CMOS cells, achieving scalability and strong resilience against de-camouflaging attacks.
Owing to these benefits, our proposed camouflage framework adopts covert gates as a core primitive.

\subsection{IP Camouflaging}
\label{sec::IPCamo}
IP Camouflage~\cite{fan2025designing} represents a paradigm shift in hardware security by introducing ``mimetic deception.''~\cite{pawlick2019game, beltran2025cyber} Unlike traditional camouflaging, which merely obscures logic, this method aims to make a specific functional circuit ($F$) structurally mimic a completely different appearance circuit ($A$). The methodology achieves this through a two-stage generative process. First, it utilizes an \textbf{And-Inverter Graph Variational Autoencoder (AIG-VAE)} to encode both $F$ and $A$ into a latent space. It then performs a linear interpolation between their latent vectors to decode a new, ``blended'' circuit topology that shares characteristics of both.

Second, and critically, this blended design undergoes a \textbf{post-generation rectification} step. Since the raw generative output is rarely functionally perfect, this phase compares the blended circuit against the target function $F$ and appearance $A$. It systematically inserts specialized \textbf{covert gates}~\cite{shakya_covert_2019}---such as Fake Inverters and Fake Buffers---to resolve discrepancies. These gates allow the circuit to physically match the visual structure of $A$ (e.g., appearing to have an inverter) while logically performing the function required by $F$ (e.g., acting as a buffer). Despite its effectiveness against SAT and GNN-RE attacks, this reliance on rectification introduces significant drawbacks:
\begin{itemize}[leftmargin=*,noitemsep]
    \item \textbf{Scale Limitation}: The AIG-VAE model is currently restricted to small logic cones of approximately 200 nodes.
    \item \textbf{Post-Generation Rectification}: The separation of generation and correction prevents a true end-to-end flow, often leading to sub-optimal mimicry.
    \item \textbf{High Overhead}: The rectification process often requires inserting additional cells to force the function-appearance mismatch, increasing Area and Power overheads.
\end{itemize}


To solve these specific problems, we propose two independent, end-to-end models that each address these flaws in different ways, presenting a trade-off of their own. We also introduce a mimicry-aware partitioning method to enhance scalability. These contributions will be detailed in the following sections. 

%% file: Section4_MimicryAwarePartitioning.tex
\section{Mimicry-Aware Partitioning} \label{sec:partition}

\begin{figure}[t]
    \centering
    \includegraphics[width=\columnwidth]{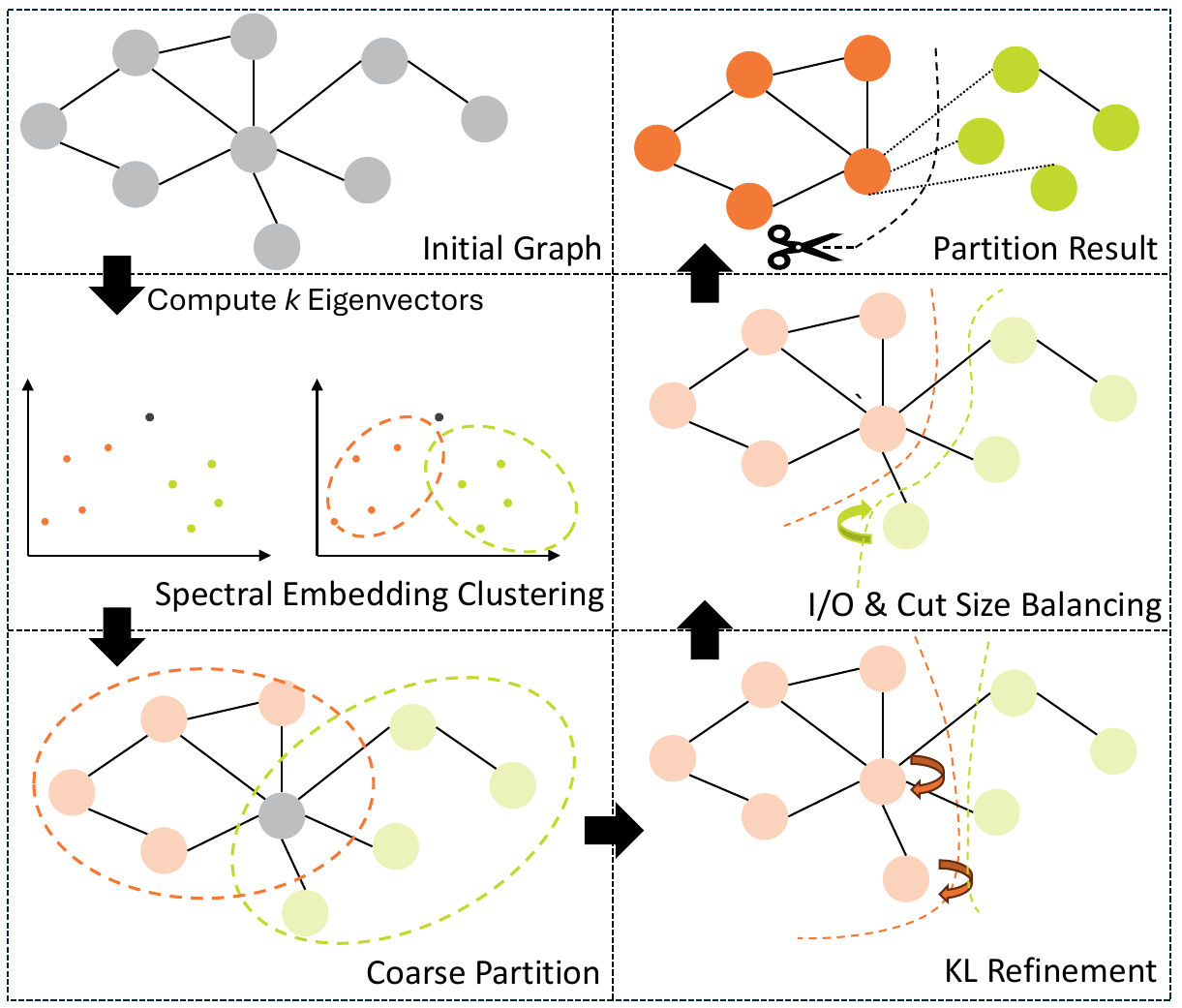}
    \vspace{-20pt}
    \caption{The three-phase hybrid partitioning flow. Coarse Partition: ``Initial Graph'' is converted into a "Spectral Embedding" to find the initial ``Coarse Partition''; KL Refinement: The partition boundary is improved to reduce the cut size; Cut Size Balancing: A greedy process moves nodes to balance I/O, producing the final ``Partition Result''.}
    \vspace{-10pt}
    \label{fig:Partition}
    
\end{figure}

To address the scalability challenge of modern, complex ICs, a common strategy is to \textbf{partition} the design into smaller, manageable subcircuits. This ``divide-and-conquer'' approach typically aims to minimize connections (the cutset) between blocks while respecting design constraints~\cite{kahng2011vlsi}. 
We implement a multi-stage heuristic that achieves \textbf{mimicry-aware partitioning} by extending these goals, optimizing for both \textbf{cutset minimization} and \textbf{I/O port balancing} per partition, subject to specified size constraints. This process consists of the following three sequential phases.


\vspace{0.5ex}

\noindent \textbf{Coarse Partitioning via Spectral Clustering.} An initial global partitioning is established using k-way Spectral Clustering~\cite{chan1993spectral}. This method is applied to the undirected adjacency matrix of the circuit graph. It groups nodes based on strong structural connectivity, effectively identifying k coarse-grained clusters (partitions) that are highly interconnected internally. The resulting total cut size is recorded as a baseline for optimization.

\noindent\textbf{Boundary Refinement via Kernighan-Lin (KL).} The second phase performs local refinement using a Kernighan-Lin (KL) bisection heuristic~\cite{dutt1993new}. It is not run on the entire graph, but instead applied iteratively to the subgraph formed by each adjacent pair of partitions identified in Phase 1. This entire refinement phase is tentative; the modified partition boundaries are accepted only if the new global cut size is strictly lower than the baseline. If no improvement is found, the results of this phase are discarded.

\noindent\textbf{Greedy I/O and Cut Size Balancing.} A greedy iterative refinement process optimizes for both cut size and I/O balance by moving individual boundary nodes to adjacent partitions. A move is selected based on an objective function that computes a weighted sum of the cut size gain and the I/O imbalance reduction. The single move with the highest positive gain is executed in each iteration. This process repeats until no moves yield a positive gain, at which point the algorithm has converged to a local optimum.

%% file: Section5_GraphMatching.tex
\section{End-to-End Methods for Deceptive Design} \label{sec:Method}

To eliminate the high-overhead \textbf{post-generation rectification} step in prior work, we introduce two novel, end-to-end methods for deceptive design. Both methods share the objective of directly synthesizing a single circuit ($G_{out}$) that is functionally equivalent to the target functional circuit ($G_F$) while being structurally similar to the appearance circuit ($G_A$). The following sections detail our two independent approaches: a standard-gate graph-matching heuristic and a DNAS-based synthesis model.

\subsection{Graph Matching Method}

\begin{algorithm}[t]
\caption{Greedy Layer-by-Layer Graph Matching}
\label{alg:graph_match_simple}
\begin{algorithmic}[1]
\State \textbf{Input:} $G_A$ (Appearance Graph), $G_F$ (Functional Graph)
\State \textbf{Output:} $M_{total}$ (Node mapping $G_A \to G_F$)

\Function{MatchGraphs}{$G_A, G_F$}
    \State $layers_A \gets \Call{LevelizeGraph}{G_A}$
    \State $layers_F \gets \Call{LevelizeGraph}{G_F}$
    \State $M_{total} \gets \emptyset$

    \State $M_{prev} \gets \Call{MatchPILayer}{layers_A[0], layers_F[0]}$
    \State $M_{total} \gets M_{total} \cup M_{prev}$

    \Comment{Iterate through all non-PI layers}
    \For{$k \gets 1$ \textbf{to} $\text{length}(layers_A) - 1$}
        \State $L_A \gets layers_A[k]$; $L_F \gets layers_F[k]$
        
        \Comment{Build cost matrix based on M\_prev}
        \State $C \gets \Call{BuildCostMatrix}{L_A, L_F, M_{prev}}$

        \State $M_k \gets \Call{HungarianAlgorithm}{C}$

        \State $M_{total} \gets M_{total} \cup M_k$
        \State $M_{prev} \gets M_k$ \Comment{Set up for next layer}
    \EndFor
    \State \textbf{return} $M_{total}$
\EndFunction
\end{algorithmic}
\end{algorithm}

To address the representation mismatch of AIG-based methods, we propose a graph-matching heuristic that operates directly on standard logic gate netlists as shown in Algorithm \autoref{alg:graph_match_simple}. Its goal is to find a consistent, optimal mapping from the \textbf{appearance graph ($G_A$)} to the \textbf{functional graph ($G_F$)}. This is achieved using a novel \textbf{layer-by-layer greedy heuristic}, where the matching established at layer $k$ becomes the fixed basis ($M_{prev}$) for calculating costs at layer $k+1$.

\noindent \textbf{Graph Levelization.}
As a pre-processing step, both DAGs, $G_A$ and $G_F$, are independently 'levelized' using a topological sort (Kahn's algorithm~\cite{kahn1962topological}). This assigns every node to a layer ($Layer_0$ being PIs), which is crucial as the matching algorithm requires strict layer-to-layer correspondence.

\noindent \textbf{Layer-by-Layer Matching.}
The core algorithm iterates from layer $k=1$ to the final layer, performing a \textbf{minimum-cost bipartite matching} at each step. This process begins by matching PIs based on names, and then follows these steps for each subsequent layer $k$:
\begin{itemize}[leftmargin=*,noitemsep]
\item \textbf{Cost Matrix Construction:} A cost matrix is built where each entry $(i, j)$ represents the cost of matching node $a_i \in Layer_k(G_A)$ to node $f_j \in Layer_k(G_F)$.
\item \textbf{Asymmetric Cost Function:} The cost is calculated as a sum of a \textbf{Node Cost} and a \textbf{Connection Cost}, designed specifically for mimicry based on the $M_{prev}$ matching from layer $k-1$.
\item \textbf{Assignment:} The \textbf{Hungarian algorithm} is used to find the optimal assignment for layer $k$.
\end{itemize}
If layers are unbalanced, dummy nodes with a high cost are used to ensure the assignment algorithm can leave some nodes unmatched.

\noindent \textbf{Asymmetric Cost Function.}
The core heuristic relies on its asymmetric cost function to evaluate the ``goodness'' of matching $a \in G_A$ to $f \in G_F$:
\begin{itemize}[leftmargin=*,noitemsep]
\item \textbf{Node Cost:} The cost of matching logic functions is explicitly defined. A low penalty (cost 0) is applied when the functional logic is smaller and can be easily hidden within the appearance logic (e.g., hiding a NOT gate in a NAND/NOR gate, hiding a BUF gate in a AND/OR gate, or hiding any gate in a XOR/XNOR gate), aligning with the capability of Covert Gate~\cite{shakya_covert_2019} technology.
\item \textbf{Connection Cost:} This cost is based on the matching from the previous layer, $M_{prev}$. It compares the set of predecessors of $f$ with the set of \textit{mapped} predecessors of $a$. The cost is increased for every connection present in $F$ that is \textit{not} present in the mapped $A$ (\texttt{F\_predecessors - mapped\_A\_predecessors}). This enforces a containment constraint, encouraging $G_F$'s connectivity to be a subset of $G_A$'s.
\end{itemize}
Based on this final mapping, the deployment of covert components is performed.

%% file: Section6_NANDGateArray.tex
\subsection{NAND Gate Array Method} \label{sec:NAND}

\begin{figure*}[t]

    \centering

    \includegraphics[width=1\textwidth]{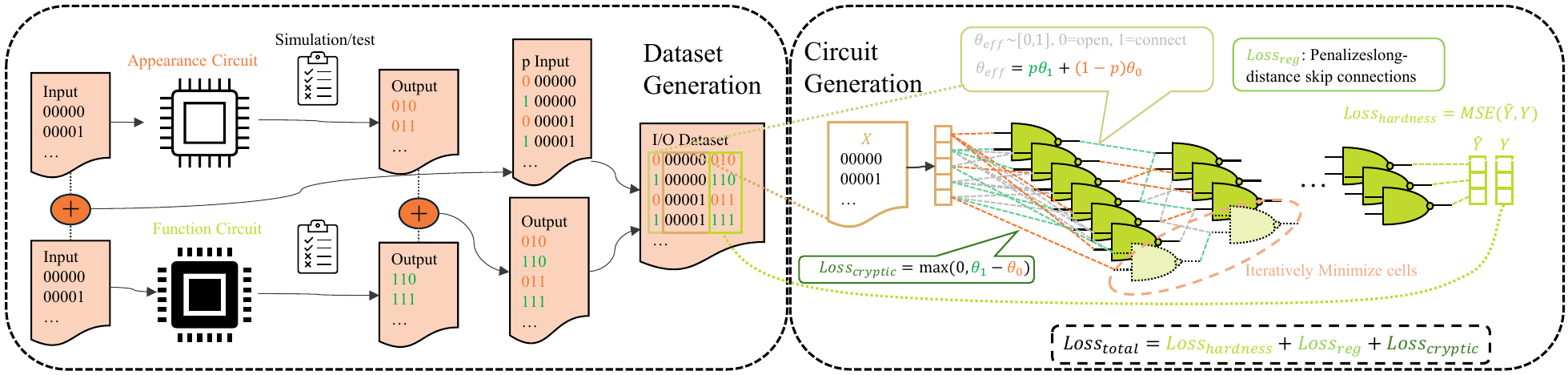}
    \vspace{-10pt}
    \caption{The DNAS-based NAND Gate Array (SelectorTNet) workflow. A unified I/O dataset is first generated from the test vectors of the \textbf{Appearance Circuit ($p=0$)} and the \textbf{Function Circuit ($p=1$)}. This dataset trains the SelectorTNet to synthesize a single NAND-gate array. The model is guided by a composite $Loss_{total}$, which includes \textbf{$Loss_{hardness}$} for functional correctness, \textbf{$Loss_{reg}$} to penalize skip connections, and \textbf{$Loss_{cryptic}$} to ensure the function's connections are a subset of the appearance's.}

    \label{fig:NAND}

\end{figure*}

Recent advances in DNAS (differentiable neural architecture search)-based logic synthesis have shown it is possible to generate a circuit directly from its input-output (I/O) examples, or truth table~\cite{belcak2022neural,wang2024towards}. This state-of-the-art approach offers a path to creating highly optimized, end-to-end circuit models. To solve the scalability and end-to-end challenge for IP mimicry, we adapt this concept by extending the T-net architecture~\cite{wang2024towards}. We call our new model \textbf{SelectorTNet}, which is designed to synthesize a single, blended NAND-gate array by learning two distinct logic functions -- our Functional Circuit ($G_F$) and Appearance Circuit ($G_A$) -- within a unified framework.

This is achieved by incorporating a selector variable $p$, which governs the model’s functional behavior. Per our design, $p=1$ selects the $G_F$ behavior, while $p=0$ selects the $G_A$ behavior. Specifically, the network maintains two distinct sets of learnable parameters for all node and output connections -- $\theta_0$ corresponding to the \textbf{Appearance Circuit ($p=0$)} and $\theta_1$ corresponding to the \textbf{Functional Circuit ($p=1$)}. During the forward pass, the selector input $p$ dynamically determines the effective parameter set $\theta_{\text{eff}}$ for each batch instance through a differentiable interpolation:
\begin{equation}
    \theta_{eff} = p \cdot \theta_1 + (1-p) \cdot \theta_0
\end{equation}
This interpolation enables a smooth, differentiable transition between the two functional regimes. This allows the network to learn a single, blended circuit that satisfies both logical behaviors, eliminating the need for the post-generation rectification from~\cite{fan2025designing}.

Structurally, SelectorTNet preserves the triangular connection topology and differentiable NAND-based computation characteristic of the original T-Net design. It employs Gumbel-Softmax sampling at each node to ensure sparse and differentiable selection of active connections, supporting end-to-end gradient-based optimization.
To effectively train the SelectorTNet for IP mimicry, we adapt the complementary loss functions from T-Net to jointly guide its learning process:

\vspace{0.5ex}

\noindent \textbf{Hardness-aware Loss.} This loss drives the model toward perfect logical accuracy (100\%) for both circuits. It is the primary objective for ensuring the final blended circuit is \textbf{functionally correct} to $G_F$ when $p=1$, and correctly matches the $G_A$ truth table when $p=0$. This is adapted from the T-Net's loss function, which penalizes ``hard'' examples to eliminate residual errors.

\vspace{0.5ex}

\noindent \textbf{Inner-Architecture Regularized Loss.} To promote efficient logic representations, this regularization term penalizes long-distance ``skip'' connections. This encourages the formation of structured, sequential computation paths, resulting in synthesis-friendly netlists.

\vspace{0.5ex}

\noindent \textbf{Cryptic Loss (Connection Containment).} This is the core loss function for achieving mimicry, adapted from the \textit{Connection Containment Loss} in T-Net. It enforces a structural containment constraint, ensuring the \textbf{Functional Circuit ($p=1$) remains a structural subset of the Appearance Circuit ($p=0$)}. Specifically, it penalizes any connection in the $G_F$ circuit ($p=1$) that exhibits a stronger activation ``desire'' than its counterpart in the $G_A$ circuit ($p=0$), using a ReLU-based formulation: $Loss_{cryptic} = \max(0, p1_{prob} - p0_{prob})$

This asymmetric penalty forces any connection required for the function to \textit{also} be present in the appearance circuit. This effectively ``hides'' the true functional circuit within the ``sea of gates'' of the larger appearance (decoy) circuit.

Together, these loss components jointly enforce the SelectorTNet’s learning objectives. The overall training objective integrates these components into a unified formulation:
\begin{equation}
    Loss = Loss_{hardness} + \lambda_{reg} \cdot Loss_{reg} + \lambda_{cryptic} \cdot Loss_{cryptic}
\end{equation}
where $\lambda_{\text{reg}}$ and $\lambda_{\text{cryptic}}$ are weighting coefficients. This composite loss enables the SelectorTNet to jointly optimize for functional accuracy and structural mimicry, producing a compact, correct, and deceptive circuit in a single end-to-end process. These coefficients for the training are set to $\lambda_{\text{reg}} = 0.15$ and $\lambda_{\text{cryptic}} = 10$. The higher value for $\lambda_{\text{cryptic}}$ explicitly prioritizes the enforcement of structural mimicry and security.

%% file: Section7_ExperimentAndAnalysis.tex
\section{Experimental Results and Analysis} \label{sec:results}

\begin{table*}[t] 
\scriptsize
\centering
\caption{Camouflage results with power, area, and delay overhead}
\vspace{-10pt}
\label{tab:result}
\resizebox{\linewidth}{!}{%
\begin{tabular}{ccccccccccccc}
\hline
\multicolumn{1}{l}{} &
  \multicolumn{1}{l}{} &
  \multicolumn{1}{l}{} &
  \multicolumn{3}{c}{\textbf{ICCAD~\cite{fan2025designing}}} &
  \multicolumn{3}{c}{\textbf{Graph Matching (Ours)}} &
  \multicolumn{4}{c}{\textbf{NAND Array (Ours)}} \\ \hline
Size &
  Functional (\#nodes) &
  Appearance (\#nodes) &
  \textbf{Area} &
  \textbf{Power} &
  \textbf{Delay} & 
  \textbf{Area} &
  \textbf{Power} &
  \textbf{Delay} &
  \textbf{Area} &
  \textbf{Power} &
  \textbf{Delay} &
  \textbf{Acc (\%)} \\ \hline
Tiny                    & c17 (13)            & mux\_4 (20)      & 1.55$\times$ & 1.47$\times$ & 1.0$\times$ & 2.92$\times$ & 4.33$\times$ & 1.60$\times$ & \textbf{1.16$\times$} & \textbf{1.26$\times$} & 1.0$\times$ & 95.3$\%$ \\ \hline
\multirow{3}{*}{Small}  & banyan\_8 (336)     & ctrl (342)       & 1.68$\times$ & 1.62$\times$ & 1.0$\times$ & 1.73$\times$ & 2.09$\times$ & 1.0$\times$  & \textbf{1.59$\times$} & \textbf{1.54$\times$} & 1.0$\times$ & 99.4$\%$   \\
                        & c1908 (615)         & c1355 (687)      & 1.90$\times$ & 1.72$\times$ & 1.0$\times$ & \textbf{1.39$\times$} & \textbf{1.38$\times$} & 1.32$\times$ & \textbf{1.39$\times$} & 1.42$\times$ & 1.0$\times$ & 99.3$\%$ \\
                        & c499 (703)          & banyan\_16 (992) & 1.43$\times$ & 1.38$\times$ & 1.0$\times$ & 1.56$\times$ & 1.47$\times$ & 1.63$\times$ & \textbf{1.18$\times$} & \textbf{1.21$\times$} & 1.0$\times$ & 100$\%$ \\ \hline
\multirow{2}{*}{Medium} & c5315 (2184)        & i2c (2490)       & 1.40$\times$ & 1.23$\times$ & 1.0$\times$ & 1.70$\times$ & 2.03$\times$ & 1.34$\times$ & \textbf{1.17$\times$} & \textbf{1.17$\times$} & 1.0$\times$ &  95.4$\%$\\
                        & c6288 (3539)        & bar (3990)       & 2.28$\times$ & 2.26$\times$ & 1.0$\times$ & 2.05$\times$ & 2.38$\times$ & 1.0$\times$  & \textbf{1.05$\times$} & \textbf{1.07$\times$} & 1.0$\times$ & 97.3$\%$ \\ \hline
Large                   & multipilier (45740) & log2 (51455)     & N/A          & N/A          & N/A         & 1.28$\times$ & 1.27$\times$ & 1.08$\times$ & \textbf{1.14$\times$} & \textbf{1.19$\times$} & 1.0$\times$ & 93.2$\%$ \\ \hline
\end{tabular}%
}
\end{table*}

\begin{table*}[]
\scriptsize
\centering
\caption{GNN-RE results comparing through Methods}
\vspace{-10pt}
\label{tab:GNNRE}
\resizebox{\linewidth}{!}{%
\begin{tabular}{ccccccccccccc}
\hline
\multicolumn{1}{l}{} &
  \multicolumn{1}{l}{} &
  \multicolumn{1}{l}{} &
  \multicolumn{3}{c}{\textbf{ICCAD~\cite{fan2025designing}}} &
  \multicolumn{3}{c}{\textbf{Graph Matching (Ours)}} &
  \multicolumn{3}{c}{\textbf{NAND Array (Ours)}} \\ \hline
Size &
  Functional (\#nodes) &
  Appearance (\#nodes) &
  \textbf{$F1_{expose}$} &
  \textbf{$F1_{mimicry}$} &
  \textbf{$Score$} &
  \textbf{$F1_{expose}$} &
  \textbf{$F1_{mimicry}$} &
  \textbf{$Score$} &
  \textbf{$F1_{expose}$} &
  \textbf{$F1_{mimicry}$} &
  \textbf{$Score$} \\ \hline
Tiny & c17 (13)            & mux\_4 (20)                         & 0.28 & 0.64 & 1.28  & 0.15 & 0.81 & 4.4   & \textbf{0.01}  & \textbf{0.98} & \textbf{97}   \\ \hline
\multirow{3}{*}{Small}  & banyan\_8 (336)     & ctrl (342)       & 0.17 & \textbf{0.73} & \textbf{3.29}  & 0.37 & 0.62 & 0.68  & \textbf{0.1}   & 0.26 & 1.6  \\
                        & c1908 (615)         & c1355 (687)      & 0.26 & 0.31 & 0.19  & 0.6  & 0.59 & -0.02 & \textbf{0.12}  & \textbf{0.86} & \textbf{6.2}  \\
                        & c499 (703)          & banyan\_16 (992) & 0.33 & 0.11 & -0.67 & 0.31 & \textbf{0.63} & 1.03  & \textbf{0.004} & 0.46 & \textbf{10.5} \\ \hline
\multirow{2}{*}{Medium} & c5315 (2184)        & i2c (2490)       & 0.45 & \textbf{0.56} & 0.24  & 0.22 & 0.42 & 0.91  & \textbf{0.11}  & 0.48 & \textbf{3.36} \\
                        & c6288 (3539)        & bar (3990)       & 0.31 & 0.67 & 1.16  & 0.3  & \textbf{0.69} & 1.3   & \textbf{0.14}  & 0.42 & \textbf{2}    \\ \hline
Large                   & multipilier (45740) & log2 (51455)     & N/A  & N/A  & N/A   & 0.33 & \textbf{0.63} & 0.91  & \textbf{0.22}  & 0.51 & \textbf{1.32} \\ \hline
\end{tabular}%
}
\end{table*}

To evaluate the effectiveness of our proposed models, we conduct a comprehensive set of experiments based on a rigorous ``Partition-Process-Recombine'' workflow. Our evaluation is designed to answer three key questions: \textit{1) How resilient are our models to state-of-the-art attacks (SAT and GNN-RE) compared to the ICCAD 2025 baseline?}; \textit{2) What are the overheads (area, power, delay) of our models?}; \textit{3) Do our models overcome the scaling limitations of prior work?}
\subsection{Experimental Setup}

\noindent \textbf{Benchmarks.} We utilize a diverse set of benchmarks classified by size to rigorously test scalability. Our dataset is constructed from three primary sources: standard circuits from the ISCAS'85 benchmark suite, arithmetic and control blocks from the EPFL Combinational Benchmark Suite~\cite{amaru2015epfl}, and custom-generated Banyan network topologies scaled to specific node counts.


\vspace{0.5ex}

\noindent \textbf{Partitioning Strategy.} To enable scalable processing for Small, Medium, and Large designs, we apply a mimicry-aware partitioning strategy. Each Function ($F$) and Appearance ($A$) circuit pair is decomposed into approximately 10 manageable sub-circuit pieces before processing.

\vspace{0.5ex}

\noindent \textbf{Compared Methods.} We compare three different methodologies: (1) \textbf{ICCAD 2025 (Baseline)}: The AIG-VAE-based "IP Camouflage" method~\cite{fan2025designing}. Pieces are processed by adding virtual outputs, applying the ICCAD method, and recombining.; (2) \textbf{Graph Matching (Ours)}: Our standard-gate graph matching method. A-F pairs are partitioned, matched layer-by-layer, and recombined; (3) \textbf{NAND Array (Ours)}: Our DNAS-based model. Pieces are partitioned, minimized via DNAS, converted to Truth Tables, synthesized into deceptive NAND arrays, and recombined. 

\vspace{0.5ex}

\noindent
\textbf{Overhead Estimation:} We evaluate PPA (Power, Performance, Area) using logic-level proxy metrics. Area is quantified by the total count of different logic nodes. Power is estimated based on the total interconnect volume, leveraging the correlation between connection quantity and dynamic power consumption~\cite{magen2004interconnect}. Delay is measured as the logical depth of the circuit's critical path.

\vspace{0.5ex}


\noindent \textbf{Camouflaging Results.} We evaluated the scalability and overheads of the proposed methods against the baseline.
\textbf{Scalability:} The ICCAD baseline fails to process large-scale designs due to inherent input size limitations. In contrast, both our Graph Matching and NAND Array methods successfully scale to industrial-sized benchmarks, validating the effectiveness of our partitioning strategy.
\textbf{Overhead:} The \textbf{NAND Array} method demonstrates the superior trade-off, significantly reducing area and power overheads compared to the baseline while maintaining near-zero delay impact. The SelectorTNet model utilized for this process was trained using PyTorch on an NVIDIA B200 GPU. The \textbf{Graph Matching} method incurs the similar overheads as ICCAD baseline due to strict structural matching constraints. However, a key trade-off remains: while the NAND Array offers superior PPA, as a test-vector-based DNAS method, it cannot theoretically guarantee 100\% logic correctness across the entire input space, whereas the Graph Matching approach preserves formal equivalence with a full representation of logic gates.

\subsection{Security Analysis}

\subsubsection{Resilience to SAT-based Attacks.}
To evaluate logical security, we conducted SAT-based attacks using the ICySAT framework~\cite{shamsi2019icysat}, modeling camouflaged components as key-programmable elements to resolve the function-appearance ambiguity. Experiments were executed on an AMD EPYC 7702 64-core server with 100 GB RAM using the Glucose solver. The attack failed to converge for all small, medium, and large benchmarks, terminating with either timeouts (24 hours) or out-of-memory errors. These results confirm that all 3 methods effectively preserve the high logical complexity necessary to thwart SAT-guided reverse engineering.

\subsubsection{Resilience to GNN-based Reverse Engineering}

GNN-RE~\cite{alrahis2021gnn} attacks attempt to defeat camouflage by analyzing the graph structure to classify the true function of nodes. We evaluate the deceptive capability using a specific training, validation, and testing protocol.

\vspace{0.5ex}

\noindent \textbf{Methodology.} We train a GNN model on a large volume of general circuit pieces. We then evaluate performance using a specific deception score based on two F1 metrics: $F1_{expose}$ (the attacker's accuracy in correctly identifying the \textit{Functional Circuit}) and $F1_{mimicry}$ (the attacker's rate of misclassifying the logic as the \textit{Appearance Circuit}).
We define the \textbf{Deception Score} as:
\begin{equation}
    Score = \frac{F1_{mimicry} - F1_{expose}}{F1_{expose}}
\end{equation}
A higher score indicates higher deceptive ability (i.e., the GNN is misled into classifying the functional logic as the appearance logic rather than its true identity).

\vspace{0.5ex}

\noindent \textbf{Results.} The results, summarized in ~\autoref{tab:GNNRE}, demonstrate the superior deceptive capability of our proposed methods. The \textbf{NAND Array (Ours)} method achieves a remarkable Deception Score, significantly outperforming the ICCAD baseline and the Graph Matching method.
Crucially, the NAND Array method reduces the $F1_{expose}$. This indicates that the DNAS-generated NAND structures successfully destroy the structural features GNNs rely on to identify the true function (High Crypticity). Simultaneously, it maintains a higher $F1_{mimicry}$, effectively tricking the attacker.
Furthermore, consistent with previous scalability findings, the ICCAD baseline fails to produce results for the ``Large'' benchmark (N/A), whereas both of our proposed methods successfully camouflage the large-scale multiplier/log2 pair, maintaining their deceptive integrity even at scale.

\subsection{Comparative Analysis}
We analyze the performance overheads, scalability, and qualitative trade-offs among the three methods. A comprehensive summary of these trade-offs is presented in \autoref{tab:comparison}. \textbf{ICCAD 2025} baseline serves as a middle ground, achieving moderate mimicry and overheads but critically suffering from poor scalability due to inherent input size limitations. \textbf{NAND Array} method demonstrates the superior performance trade-off, achieving the highest mimicry score and ultra-low PPA overheads while remaining fully scalable. However, as a DNAS-based approach, it faces the inherent limitation of not theoretically guaranteeing 100\% formal logic correctness. 
\textbf{Graph Matching} method offers the superior logic representation by utilizing standard gates and good scalability, but incurs similar overheads to the ICCAD baseline due to its strict structural matching constraints. 

\begin{table}[]
\centering
\caption{Comparative Analysis of Deceptive Methodologies}
\vspace{-10pt}
\label{tab:comparison}
\resizebox{\columnwidth}{!}{%
\begin{tabular}{@{}lcccccc@{}}
\toprule
\textbf{Method} & \textbf{Scalable?} & \textbf{End-to-End?} & \textbf{Overhead} & \textbf{Gate Representation} & \textbf{Mimicry} & \textbf{Logic Correctness} \\ 
\midrule
\textbf{Graph Matching (Ours)} & Yes & Yes & Moderate & Standard (Good) & Moderate & 100$\%$ \\ 
\addlinespace
\textbf{NAND Array (Ours)} & Limited & Yes & Ultra-low & NAND-only (Limited) & High & around 95$\%$ \\ 
\addlinespace
\textbf{ICCAD 2025 (Baseline)~\cite{fan2025designing}} & No & No & Moderate & AIG (Limited) & Moderate & 100$\%$ \\ 
\bottomrule
\end{tabular}%
}
\vspace{-12pt}
\end{table}

%% file: Section8_Conclusion.tex
\section{Conclusion} \label{sec:conclusion}

This paper introduces a holistic methodology that resolves the core limitations of prior ``mimetic deception'' techniques. We successfully addressed the initial scalability challenge by implementing a mimicry-aware partitioning method, providing the necessary divide-and-conquer strategy for large designs. We resolved architectural flaws by proposing two novel, end-to-end models: the Graph Matching algorithm (solving the representation mismatch by operating on standard logic gates) and the DNAS NAND Array model (eliminating the post-generation rectification flaw with a scalable, single-stage synthesis flow). Our comprehensive analysis confirms that the synergy between these innovations establishes a viable, highly resilient framework for next-generation end-to-end deceptive IP design.